\documentstyle[preprint,revtex]{aps}

\newcommand{\intd}{{\rm d}}

\newcommand{\mod}{{\rm mod}}
\newcommand{\sgn}{{\rm sgn}}
\newcommand{\Si}{{\rm Si}}

\begin{document}
\draft

\begin{title}
Test of Conformal Invariance in One-Dimensional Quantum Liquid \\
with Long-Range Interaction
\end{title}

\author{Rudolf A.\ R\"{o}mer and Bill Sutherland}

\begin{instit}
Physics Department, University of Utah, Salt Lake City, UT 84112
\end{instit}

\receipt{}

\begin{abstract}
We numerically study the momentum distribution of one-dimensional
Bose and Fermi systems with long-range interaction $g/r^2$ for the
``special'' values $g= -\frac{1}{2}, 0, 4$, singled out by random
matrix theory.
The critical exponents are shown to be independent of density and
in excellent agreement with estimates obtained from $c=1$ conformal
finite-size scaling analysis.
\end{abstract}


\tightenlines

%
%

\section{Introduction}
\label{sec-intro}

In one-dimensional (1D) quantum systems it has been argued that the
existence of a critical point at zero temperature implies asymptotic
conformal invariance at large distances \cite{p70,bpz84}.
This invariance powerfully constrains the possible behavior of a given
critical theory, parametrizing universality classes by the
{\em conformal anomaly} $c$, the value of the central charge
of the underlying Virasoro algebra \cite{bpz84}.
For $c<1$, it has been shown by Friedan, Qui and Shenker that unitarity
(reflection positivity) restricts
those values of $c$ to be quantized \cite{fqs84}.
The critical exponents are given
by the Kac-formula and the correlation functions can be determined.
Realizations of this scheme are the Ising ($c=\frac{1}{2}$),
tricritical Ising ($c=\frac{7}{10}$), three-state Potts ($c=\frac{4}{5}$),
tricritical Potts ($c=\frac{6}{7}$), and other models \cite{c87,do84}.
In addition, a complete classification can also be given for $c\geq 1$
if a continuous symmetry is assumed; e.g.\ for $U(1)$ we get the
Gaussian model ($c=1$), describing a variety of models such as the
Coulomb gas, XY-model, $s=\frac{1}{2}$ antiferromagnet and more
\cite{c87,ca87}.

The basic assumptions leading to conformal invariance of a model
may be summarized as scale invariance, rotational and translational
invariance and short-range interactions.
If any of the above properties do not hold, the conformal symmetry
will in general not be obeyed either.
In addition, it should be noted, that there is no rigorous proof of
conformal invariance at a critical point, just as for scale invariance.
Instead conformal invariance should be regarded as a principle, whose
consequences can then be compared with exact and numerical results
on specific models \cite{c87}.
In the case of long-range interactions, an explicit counterexample
may be given:
Thouless \cite{t69} showed that certain long-ranged 1D classical systems
may have a critical point. Now, since any general coordinate transformation
in one dimension is conformal, this would imply that all correlation
functions are constant, which is certainly not true.

Recently, however, a number of groups have independently applied conformal
arguments to long-range models such as the periodic $g/r^{2}$
models \cite{ha91,ky91,ko91,mz91}.
They were able to compute the conformal anomaly and the conformal scaling
dimensions, as well as various correlation exponents for Bose and Fermi
statistics as functions of the interaction strength $g$. Their predictions
agree with the exact result for the case of $g=0$, that is, free fermions
and hard core bosons.
For less trivial interaction strength, no independent verification of
these results exists so far.

The periodic $g/r^{2}$ model has been solved by means of the Bethe Ansatz
by one of us several years ago \cite{su7x}.
As it turns out, for the three ``special'' values $g= -\frac{1}{2}, 0, 4$
of the interaction strength, the ground state wave
function of this model can be viewed as a distribution function for the
eigenvalues of a random matrix from an appropriate ensemble.
Application of theorems from random matrix theory then allows the derivation
of explicit formulae for the ground state correlation functions \cite{su92}.

In this paper, we will present a high-precision study of the critical
exponents for the momentum distribution, which we will then compare to
the results obtained by the conformal conjecture.
In section \ref{sec-deriv}, we will outline the arguments of conformal
invariance leading to the predictions for the critical exponents.
Section \ref{sec-corr} introduces the $g/r^{2}$ model and its
ground state correlations in more detail.
We present the results of our numerical study in section \ref{sec-res},
where in addition we present an unusual scaling law.

%
%

\section{Derivation of Critical Exponents}
\label{sec-deriv}

Finite-size scaling analysis of classical two-dimensional
statistical systems at criticality may be used to obtain
the value of the conformal anomaly $c$ and the scaling dimensions $x_n$.
For short-range interactions that give rise to a linear dispersion
relation, i.e.\ gapless excitations, the following formulae
hold \cite{c86}: for a strip of width $L \rightarrow \infty$
and periodic boundary conditions,
\begin{eqnarray}
E_0       &\sim &\epsilon_0 L - \frac{\pi}{6} \frac{c v}{L},
\label{eqn-bfca} \\
E_n - E_0 &\sim &2 \pi \frac{x_n v}{L}.
\label{eqn-bfsd}
\end{eqnarray}
Here $E_n$ is the eigenvalue of the Hamiltonian; $E_0$ is the ground
state energy; $\epsilon_0$ is the ground
state energy density in the thermodynamic limit, and $v$ is the
velocity of the elementary excitations.
Since the partition function of a classical system of finite width
and periodic boundary conditions may be viewed as the Feynman path
integral for an infinitely long quantum chain at finite temperature,
these results apply to 1D quantum systems as well \cite{a86,bcn86}.

We now wish to evaluate the scaling dimensions for the 1D Bose and
Fermi systems interacting via the long-range interaction of the
periodic $g/r^2$ model.
This model has been solved by the asymptotic Bethe Ansatz
in \cite{su7x}, and it has been shown that the energy spectrum
may be written as
\begin{equation}
E_{\{I_j\}}
 = \sum_{j=1}^{M} \left[ \frac{2\pi}{L} I_j
   + \frac{\pi\lambda}{L} (2j - M -1) \right]^2,
\label{eqn-e}
\end{equation}
where $\{I_j\}$ denotes the set of ordered (free boson)
quantum numbers of a particular
state; $M$ is the number of particles; $L$ is the length of the
ring and $2\lambda(\lambda-1)=g$. The ground state energy $E_0$
corresponds to the state with quantum numbers $I_j=0$ for
all $j=1,\ldots,M$.
In addition, the model has a sound type excitation spectrum, and
for $\lambda=\frac{1}{2},1,2$ all correlation functions may be
calculated. We will elaborate on this last fact in more detail
in the next section.

Let us now calculate the finite-size corrections to the excited
states.
We start with the Bose system first.
An excitation corresponding to a momentum transfer $P=0$
is given by changing the number of particles $M$ to $M+ \Delta M$.
By (\ref{eqn-e}), the excitation energy is given as
\begin{eqnarray}
E(M+\Delta M) - E(M)
&= &\frac{\pi^2 \lambda^2}{L^2}
    \left\{
     \sum_{j=1}^{M+\Delta M}
      \left[ 2j - (M+\Delta M) -1 \right]^2
    -\sum_{j=1}^{M}
      \left[ 2j - M -1 \right]^2
    \right\} \nonumber \\
&\simeq
   & \pi^2 \lambda^2 d^2 \Delta M
	+ \frac{1}{L} \pi^2 \lambda^2 d \Delta M^2 \nonumber \\
&= & \mu \Delta M + \frac{2\pi v}{L} \frac{\lambda}{4} \Delta M^2
\label{eqn-deltaM}
\end{eqnarray}
Here we used the formulae for the chemical potential
$\mu=\pi^2 \lambda^2 d^2$ and the sound velocity
$v=2\pi \lambda d$, as obtained in \cite{su7x}.

Let us define the Fermi momentum as
$k_F= \pi d$. Then for Fermi {\em and}\/ Bose systems and fixed
particle number $M$, taking
$\Delta D$ particles from the ``left'' Fermi momentum to the ``right''
Fermi momentum --- or equivalently, by translating the filled Fermi
sea in momentum space --- results in another type of low-lying
excitation, i.e.\
\begin{eqnarray}
E(M;\Delta D)-E(M;0)
&= &\sum_{j=1}^{M}
    \left\{
    \left[
     \frac{2\pi}{L} \Delta D +
     \frac{\pi\lambda}{L}(2j-M-1)
    \right]^{2}
    -
    \left[
     \frac{\pi\lambda}{L}(2j-M-1)
    \right]^{2}
    \right\} \nonumber \\
&= &\frac{2\pi v}{L} \frac{1}{\lambda} \Delta D^2
\label{eqn-deltaD}
\end{eqnarray}
The momentum of this excitation is $P= 2 k_F \Delta D$.

For fermions, the same reasoning applies.
However, the quantum numbers $I_j$ are restricted to integers
now and we have to follow the selection rule
\begin{equation}
\Delta D = \frac{\Delta M}{2} (\mod\ 1).
\label{eqn-sr}
\end{equation}

By combining (\ref{eqn-deltaM}) and (\ref{eqn-deltaD}), we obtain
the scaling dimensions for the above excitations as
\begin{equation}
x(\Delta M; \Delta D) = \frac{\lambda}{4} \Delta M^2
                       +\frac{1}{\lambda} \Delta D^2.
\label{eqn-sdx}
\end{equation}
This expression is valid for both Bose and Fermi statistics
modulo the selection rule (\ref{eqn-sr}).

We have not calculated the effect of particle-hole excitations on
$x$. Although we do not need the resulting additional terms for
the calculation of the long-distance behavior of the correlations,
we remark, however, that in
order to correctly read off the right and left conformal weights
that characterize the underlying Virasoro algebra, they cannot be
ignored \cite{bpz84,c86,mz91,ky91}.

Before we continue with our derivation of the critical exponents
of the momentum distribution, let us pause and first calculate
the critical anomaly $c$ for the present model. From (\ref{eqn-e})
it follows that
\begin{equation}
E_0 \equiv E_{\{0\}} = \epsilon_0 L - \frac{1}{3} \pi^2 \lambda^2 d / L.
\end{equation}
Using the above expression for the sound velocity $v$,
we may rewrite the $L^{-1}$ correction
term as
$
- \frac{\pi v}{6 L} \lambda.
$
Comparison with (\ref{eqn-bfsd}) yields $c=\lambda$.
This is a peculiar result in as much as the critical anomaly
varies continuously with the interaction strength $\lambda$.
We could instead use the low-temperature expansion of the free energy as
given in \cite{su7x} and the formula for the critical anomaly as
in \cite{a86,bcn86}, giving
\begin{equation}
F(T) \simeq F(T=0) - \frac{\pi T^2}{6 v}
\end{equation}
and therefore $c=1$. Thus the two expressions for $c$ are in contradiction.
We will discuss this point in more detail in section \ref{sec-disc}.

Let us finally investigate the long-distance behavior of the
correlation functions.
Let the pair $(\Delta M; \Delta D)$ describe an excitation
where the particle number is changed by $\Delta M$, and
$\Delta D$ particles are transported round the ring.
If we then denote by $\phi^\dagger$ ($\phi$) the creation
(annihilation) operator of this excitation, the corresponding
leading term of the equal-time correlator has the form
\begin{displaymath}
\langle \phi^\dagger(r) \phi(0) \rangle
 \propto \cos(P r)\ r^{- 2 x(\Delta M;\Delta D)},
\end{displaymath}
where $P$ is the associated momentum carried by the excitation \cite{c87}.
Therefore the correlation of the Bose field operator has the
form
\begin{equation}
\langle \phi_B^{\dagger}(r) \phi_B(0) \rangle_\lambda
\propto r^{-\beta_{B,\lambda}},
\label{eqn-bb}
\end{equation}
where the corresponding excitation is given by
$(1;0)$, and consequently by (\ref{eqn-sdx})
$\beta_{B,\lambda}= \lambda / 2$.
Due to the selection rule (\ref{eqn-sr}) for fermions,
the corresponding excitation of the correlation
\begin{equation}
\langle \phi_F^{\dagger}(r) \phi_F(0) \rangle_\lambda
\propto \cos(k_F r)\ r^{-\beta_{F,\lambda}}
\label{eqn-bf}
\end{equation}
is given as $(1,\case{1}/{2})$ and thus
$\beta_{F,\lambda}= (\lambda+1/\lambda) / 2$.

Taking the Fourier transform of the long-distance expressions
(\ref{eqn-bb}) and (\ref{eqn-bf}), we get the momentum distribution
function $n(k)$.
The long-distance behavior now translates into the behavior at
long-wavelength or small wave numbers, and thus for the Bose
momentum distribution near the origin,
\begin{equation}
n_B(k)\propto |k|^{\alpha_{B,\lambda}}
       =       |k|^{\beta_{B,\lambda}-1}.
\label{eqn-ab}
\end{equation}
For the Fermi case, we have correspondingly for the momentum
distribution near $k_F$,
\begin{equation}
n_F(k)\propto |k-k_F|^{\alpha_{F,\lambda}} \sgn(k-k_F)
       =       |k-k_F|^{\beta_{F,\lambda}-1} \sgn(k-k_F)
\label{eqn-af}
\end{equation}
In Fig.\ \ref{fig-ce} we show the behavior of the critical exponents
as functions of $\lambda$. We emphasize that the exponents
do not depend on the particle density $d$, but on $\lambda$
only.

Our above derivation of the critical exponents follows largely the
article of Kawakami and Yang \cite{ky91}.
Other derivations have been given by Mironov and Zabrodin, who obtained
the correlation exponents for a wider class of 1D quantum models with pairwise
interaction \cite{mz91}, and by Kolomeisky \cite{ko91}, who applied methods
of two-dimensional elasticity theory to arrive at corresponding results.

%
%

\section{Correlations in the Sutherland Model}
\label{sec-corr}

The ground state wave function for the periodic $g/r^2$ model
with $M$ particles, either bosons or fermions, interacting on a
one-dimensional ring of circumference of length $L$,
is of the Jastrow product form and given by \cite{su7x,su92}
\begin{eqnarray*}
\Psi_\lambda &=
&\left(
  \frac{\Gamma(1+\lambda)^M}{L^M \Gamma(1+\lambda M)}
 \right)^{\frac{1}{2}}
 \prod_{1\leq j<k\leq M}
  | 2 \sin\left[ \pi(x_k-x_j)/L \right] |^\lambda S \\
&\equiv
&C_{\lambda,M}
 \prod_{1\leq j<k\leq M}
  \psi_\lambda\left[ 2\pi(x_k-x_j)/L \right].
\end{eqnarray*}
The factor $S$ is equal to $1$ for bosons, while for fermions
$S=(-1)^P$ is the parity of the particle ordering permutation $P$,
$x_{P1}<\cdots <x_{PM}$.
The periodicity in $L$ is assured if $M$ is odd.
$\psi_\lambda(\theta) = |2 \sin(\theta/2)|^{\lambda}$ for bosons
and $\psi_\lambda(\theta) = |2 \sin(\theta/2)|^{\lambda} \sgn(\theta)$
for fermions, the normalization constant is given as
\begin{displaymath}
C_{\lambda,M}^2 = \frac{\Gamma(1+\lambda)^M}{L^M \Gamma(1+\lambda M)}.
\end{displaymath}

As has been noted in \cite{su7x}, the square of $\Psi_\lambda$ for
the ``special'' values $g= -\frac{1}{2}, 0, 4$
--- or $\lambda= \frac{1}{2}, 1, 2$ ---
may be recognized as being identical to the joint probability
density function for the eigenvalues of matrices from
Dyson's circular ensemble \cite{d62},
$\lambda= \frac{1}{2}, 1$ and $2$ corresponding to
orthogonal, unitary and symplectic ensembles, respectively.

Results from the theory of random matrices now enable the calculation
of various correlation functions \cite{m}. We will use the following
result: For $u(\theta) = u(-\theta)= u(2\pi + \theta)$, arbitrary
otherwise, and the special values of $\lambda$,
\begin{equation}
\langle
 \prod_{j=1}^{M} u(\theta_j)
\rangle_{\lambda}
= \det[ F_{pq} ],
\label{eqn-detf}
\end{equation}
where for $\lambda=1$,
\begin{displaymath}
F_{pq}=
 \frac{1}{2\pi} \int_{-\pi}^{\pi} \intd{\theta} u(\theta) \cos(p-q)\theta,
\end{displaymath}
\begin{displaymath}
p,q = \case{1}/{2}, \case{3}/{2}, \ldots, M-\case{1}/{2};
\end{displaymath}
for $\lambda=2$,
\begin{displaymath}
F_{pq}=
 \frac{1}{2\pi p} \int_{-\pi}^{\pi} \intd{\theta} u(\theta)
  \left[ p \cos(p\theta)\cos(q\theta) + q \sin(p\theta)\sin(q\theta) \right],
\end{displaymath}
\begin{displaymath}
p,q = \case{1}/{2}, \case{3}/{2}, \ldots, M-\case{1}/{2};
\end{displaymath}
finally for $\lambda=\frac{1}{2}$,
\begin{displaymath}
F_{pq}=
 \frac{p}{4\pi} \int_{-\pi}^{\pi} \intd{\theta} u(\theta)
                \int_{-\pi}^{\pi} \intd{\phi} u(\phi) u(\phi)
  		\sgn(\theta-\phi)
  \left[ \cos(p\phi)\sin(q\theta) - \cos(p\theta)\sin(q\phi) \right],
\end{displaymath}
\begin{displaymath}
p,q = \case{1}/{2}, \case{3}/{2}, \ldots, \case{M}/{2}-\case{1}/{2}
\end{displaymath}
and we assume $M$ to be even for all three cases.
Thus, $F_{pq}$ is an $M\times M$ matrix for $\lambda= 1,2$ and an
$\frac{M}{2}\times\frac{M}{2}$ matrix for $\lambda=\frac{1}{2}$.

The one-particle density matrix is defined as
\begin{displaymath}
\rho(x-x') =
 M \prod_{j=1}^{M-1}
    \int_{0}^{L} \intd{x_j}
	\Psi_{\lambda}(x_1,\ldots,x_{M-1},x)
	\Psi_{\lambda}(x_1,\ldots,x_{M-1},x'),
\end{displaymath}
which with the help of (\ref{eqn-detf}) may be written as
\begin{eqnarray}
\rho(x-x') &= &
 M \frac{C_{\lambda,M}^2}{C_{\lambda,M-1}^2}
 \langle
  \prod_{j=1}^{M-1}
   \psi_{\lambda}(\theta_j - 2\pi x/L)
   \psi_{\lambda}(\theta_j - 2\pi x'/L)
 \rangle_{\lambda,M-1} \nonumber \\
&= &
 \frac{M}{L}
 \frac{\Gamma(1+\lambda) \Gamma(1 + \lambda(M-1))}
      {\Gamma(1+\lambda M)}
 \det[ F_{pq} ].
\label{eqn-rho}
\end{eqnarray}
In the last formula, we use the appropriate matrix $F_{pq}$ with
$M$ replaced by $M-1$. By the periodicity condition on the wave
function, we restrict $M$ to be odd. Thus, $M-1$ is even and our
above formulae are consistent.
The normalization of $\rho$ has been chosen such that
$\rho(0) = M/L = d$.
The momentum distribution $n(k)$ is defined as the Fourier transform
of $\rho$, i.e.\
\begin{displaymath}
 n(k) = \int_{-\infty}^{\infty} \intd{r} \rho(r) e^{- i k r}
\end{displaymath}
and the normalization is such that
\begin{displaymath}
 2\pi d = \int_{-\infty}^{\infty} \intd{k}\: n(k).
\end{displaymath}
The $\lambda=1$ case corresponds to free fermions or hard core bosons.
Free fermions are well understood, and we find
\begin{displaymath}
\rho(r) = \frac{\sin(\pi d r)}{\pi r}, \qquad
n(k) = \left\{
 \begin{array}{ll}
   \frac{1}{2\pi}, &|k| < \pi d \\
   0,              &|k| > \pi d
 \end{array}\right.
\end{displaymath}
Thus, $\alpha_{F,1} = 0$ in agreement with (\ref{eqn-af}).
Hard core bosons have been treated extensively in the literature
\cite{gi60,sc63,le64}. There is no condensation into a single momentum state,
and it is proven that $n(k)$ diverges as $|k|^{-\frac{1}{2}}$
at the origin. Again, this is in agreement with the corresponding result of
section \ref{sec-deriv}.

For $\lambda=2$, the matrix $F_{pq}$ has a particularly simple
Toeplitz form, which allows the explicit calculation of $\rho$ \cite{su7x},
\begin{displaymath}
 \rho(r) = \frac{\Si(2\pi d r)}{2\pi r}, \qquad
 \Si(r)\equiv\int_{0}^{r} \intd{x} \frac{\sin(x)}{x}
\end{displaymath}
\begin{eqnarray}
 n(k) = \left\{
 \begin{array}{ll}
  \frac{\ln(2\pi d /|k|)}{4\pi}, &|k| \leq 2\pi d \\
  0,                             &|k| \geq 2\pi d
 \end{array}\right.
\label{eqn-nkb2}
\end{eqnarray}
Thus, the conformal result $\alpha_{B,2} = 0$ agrees again; although,
the logarithmic behavior is not universal and so does not follow from
the finite-size analysis.

This leaves the remaining three cases ---
bosons with $\lambda=\frac{1}{2}$,
and fermions with $\lambda=\frac{1}{2}$ and $\lambda=2$
--- to be investigated numerically.

Let us now study the restriction of the above family of wave functions
to a lattice. The size of the systems $L$ is now an integer $N$, and
the coordinates of the particles are restricted to integers
$x_j= 1, 2, \ldots, N$.
The wave functions are nearly as before; the only substantial
change is the normalization constants $C_{\lambda,M}$.
For the special values of $\lambda$, the new $C'_{\lambda,M}$ are
known \cite{su92}.
The one-particle density matrix $\rho(r)$ and the corresponding
momentum distribution function $n(k)$ for the lattice case may
still be calculated as before, if we replace the integrals by
appropriate sums over lattice sites.

Note, however, that whereas the density $d=M/L$ in the continuum case
entered trivially into all expressions as a length scale, the new
density $d'=M/N$ in the discrete case is dimensionless and will
thus enter all expressions in an essential way.
Indeed, the previous continuum results are the zero-density limit of
the discrete case.
We showed in section \ref{sec-deriv} that the critical exponents
of the momentum distribution, as obtained from conformal analysis, should
be independent of density.
This prediction may therefore be verified by studying both continuum and
lattice cases.

However, we want to emphasize that whereas the wave functions for the
continuous case describe the ground state of the periodic $g/r^2$ model,
there need not be any corresponding Hamiltonian for the wave functions
of the lattice gas. For $\lambda=1$, we of course have either hard
core bosons of free fermions; while for $\lambda=2$ bosons, we have
the Haldane-Shastry model \cite{gv87,s88,h88} of a Heisenberg magnet
with $1/r^{2}$ exchange.
However, for the remaining three cases --- precisely the cases we
are to evaluate numerically --- it is not clear to what specific model
our results belong to in the discrete case. From the conformal viewpoint,
of course, the results represent a universality class of models.

Finally, the lattice gas can be shown to exhibit a particle-hole
symmetry, such that we only need to consider $d\leq\frac{1}{2}$.
In addition, at half-filling, Fermi statistics ensure
$\rho(r)=0$ for $r\neq 0$ or even and therefore for the momentum distribution,
the symmetry $n(k) + n(k+\pi) = 1$ holds.

%
%

\section{Results}
\label{sec-res}

Using the results obtained in section \ref{sec-corr}, we calculate the
one-particle density matrix $\rho(x-x')$, or by translation invariance
$\rho(r)$, with the help of (\ref{eqn-rho}) in terms of
the appropriate matrix $F_{pq}$.
Taking the Fourier transform, we compute the momentum distribution
$n(k)$, which, for fixed $L$ and $N$, turns out to have a slightly
different quantitative behavior for the continuum and discrete cases.
In Fig.\ \ref{fig-nkb} and \ref{fig-nkf}, we show $n(k)$ corresponding
to the discrete case at the special values $\lambda= \frac{1}{2},1,2$
and half-filling, for bosons and fermions, respectively.

The extraction of the critical exponents $\alpha_{B,\lambda}$ for
Bose and $\alpha_{F,\lambda}$ for Fermi statistics
is complicated by finite-size effects. The interchange of
the $L\rightarrow\infty$ and $k\rightarrow 0$ ($k\rightarrow k_F$)
limits is non-trivial, and the
correct order --- $L\rightarrow\infty$ limit first --- is hard to achieve.
In addition, we know from (\ref{eqn-nkb2}) that the momentum
distribution for the bosonic $\lambda=2$ case has a logarithmic
behavior, which is not predicted by the conformal analysis, and
the possible appearance of this or some other finite-size effect
for the remaining cases may further obscure the extraction
of the correlation exponents.

However, a careful analysis of our data reveals that the
one-particle density matrix $\rho$ obeys an unusual scaling
law. For different lattice sizes $L$, a limiting curve
$\rho_\infty$ may be constructed as
\begin{equation}
\rho_\infty(r)= \frac{1}{L^{\beta_{\lambda}}} \rho_L(r/L),
\label{eqn-sc}
\end{equation}
both for Bose and Fermi statistics.
Therefore, the limit $L\rightarrow \infty$ of $\rho_L(r)$ may be
taken at any point $r$,
and we can directly extract the asymptotic value of the
critical exponents of the momentum distribution by studying the
long-distance behavior, i.e.\ $\rho(L/2)$.
In Fig.~\ref{fig-lim} we show the remarkably good convergence of
our estimates for the critical exponents as function of the inverse
length $1/L$ of the ring for the two $\lambda=\frac{1}{2}$ cases and
the fermionic $\lambda=2$ case.

The scaled one-particle density function exhibits another quite
remarkable property. In Fig.'s \ref{fig-rhob} and \ref{fig-rhof},
we have plotted $\rho_{\infty}$ for bosons and fermions and
normalized it so that its value at the end points equals $1$.
Although, as we remarked above, $n(k)$ is different for the discrete
and the continuous case, $\rho_{\infty}$ is identical for both
cases up to the accuracy of our calculation and thus independent of
density.
In addition, the curves corresponding to $\lambda=\frac{1}{2}$ and
$\lambda=2$ in Fig.~\ref{fig-rhof} lie on top of each other.
This seems to indicate that $\rho_{\infty}$ is the same for
identical values of the critical exponent $\alpha$.
This is further supported by the fact that the bosonic $\lambda=2$
curve, corresponding to $\alpha=0$, is the same as the fermionic
$\lambda=1$ case, corresponding to $\alpha=0$, too.
We will discuss the implications of these findings in more detail
in the next section.

We have calculated the critical exponents, both for the bosonic and
fermionic cases and $\lambda= 1,2$, for up to $M=641$
particles on $L=1282$ sites. Since the scaling curves are identical
for discrete and continuum cases, so are the exponents.
An independent estimate from $n(k)$ directly confirms this.
For $\lambda=\frac{1}{2}$, the continuum formulae require
the computation of double integrals of transcendental functions
rather than orthogonal functions, and so
do not reduce to the calculation of simple Kronecker
$\delta$'s. Therefore, calculations could be done only
up to $M=41$ particles at half-filling.
The discrete expressions require less computational effort
and data for up to $M=161$ for the bosonic case, and
$M=321$ for the fermion case have been calculated.
Again, the results indicate that the critical exponents are
the same, independent of whether the continuum or the discrete
formulae are used.
Since the discrete case is the low-density limit of the
continuous case, we have therefore established that up
to our numerical accuracy, the critical exponents
$\alpha_{F,\lambda}$ and $\alpha_{B,\lambda}$ are
independent of the density $d$ as predicted by the conformal
arguments.

Let us now compare the values of our numerical estimates
with the predictions of section \ref{sec-deriv} for the
critical exponents.
Preliminary estimates for small particle numbers indicated a
disagreement with the predictions of the conformal analysis
for the bosonic $\lambda=\frac{1}{2}$ case \cite{su92}.
In table \ref{tab-res}, we show that our present study shows
excellent agreement of conformal and numerical results;
the estimates for the exactly known results are included as
a check on the numerical analysis.

Finally, we note that we could not find any logarithmic
behavior except --- as predicted --- for the bosonic $\lambda=2$
case. However, there do seem to be substantial finite-size effects.

%
%

\section{Discussion}
\label{sec-disc}

In the preceding sections, we have reviewed arguments of conformal
finite-size scaling analysis leading to predictions for the critical exponents
of the momentum distribution in the 1D quantum liquid with long-range
interaction $g/r^2$.
Both for bosonic and fermionic systems and at the special values
$g=-\frac{1}{2},0,4$ for the interaction strength, we then proceeded
to calculate the same exponents by use of explicit formulae as given by
random matrix theory.
For non-trivial cases, no analytic solutions exist and we calculated
the exponents numerically. This is most conveniently done by discretizing
the continuous problem to a lattice problem.
The continuum results are then recovered by taking the low-density limit
of the lattice system.
Our results show that the values of the correlation exponents are
the same for lattice and continuum case, thus independent of density.
This verifies the corresponding prediction of conformal field theory.
In addition, the numerical estimates of $\alpha_B,\alpha_F$ are in
excellent agreement with the conformal predictions.

The numerical analysis was helped by the appearance of the unusual
scaling law (\ref{eqn-sc}). As it turns out, the scaled and
properly normalized one-particle density function
$\rho_{\infty}$ may be parametrized by the critical exponent
$\alpha$ only, independently of statistics and interaction strength
$\lambda$. Furthermore, the interacting Bose gas at $\lambda=2$
has the same scaled $\rho_{\infty}$ as the non-interacting free
Fermi gas. This seems to suggest that not only the strength of the
interaction but also the specific type of the interaction is irrelevant.
We do not know whether this behavior may be explained in terms of
the conformal symmetry of the system, but hope to come back to
this point in a future publication.

Trying to classify the $g/r^2$ model by its critical anomaly, we
calculated the finite-size corrections to the ground state energy as
in (\ref{eqn-bfca}).
We found that in the cylindrical geometry, $c$ depends continuously on
the value of the coupling constant $\lambda$.
On the other hand, use of the low-temperature expansion of the
free energy showed that $c=1$.
Furthermore, the exponents
$\alpha_B$, $\alpha_F$ vary continuously as functions of $\lambda$.
In \cite{ca87} this has been shown to be a typical phenomena of
conformal field theories with anomaly $c=1$.
In addition, Kawakami and Yang \cite{ky91} demonstrated that
$\beta_B$, $\beta_F$ in our model obey the scaling relations of the
$c=1$ Luttinger liquid.
Therefore, we believe that $c=1$ is the correct value of the critical
anomaly of the $g/r^2$ model.
We emphasize that for models with $c\geq 1$ there is no Kac-formula,
relating the critical anomaly and the scaling dimensions, so we
proceeded to calculate the correlation exponents independently of $c$.

We conclude this paper by noting that at present, we do not
know a satisfactory rigorous explanation for the observed discrepancy
in the estimate of $c$ in terms of conformal field theory.
However, we believe this to be a remainder of
the long-range character of the $g/r^2$ interaction, and
in the light of our introductory remarks of section \ref{sec-intro},
it should not be too surprising if time correlations
differ from space correlations in long-ranged models.

\acknowledgements
R.A.R.\ would like to thank Carlton De Tar for helpful comments.


\figure{Correlation exponents for bosonic and fermionic systems
as functions of the interaction strength $g$. The values
corresponding to the special points $g=-\frac{1}{2},0,4$ are indicated.
\label{fig-ce}}

\figure{Momentum distributions for the bosonic system
on a lattice at half-filling, for the special values of $\lambda$.
\label{fig-nkb}}

\figure{Momentum distributions for the fermionic system
on a lattice at half-filling, for the special values of $\lambda$.
\label{fig-nkf}}

\figure{Estimates of the critical exponents as functions of the
inverse length of the lattice.
($\Box$) corresponds to $\beta_{B,\frac{1}{2}}$,
($+$) to $\alpha_{F,2}$ and
($\Diamond$) to $\alpha_{F,\frac{1}{2}}$.
\label{fig-lim}}

\figure{Scaled one-particle density $\rho_\infty$ for the bosonic system
at half-filling, for the special values of $\lambda$. The curves for the
discrete and the continuous case are identical up to the accuracy of the
plot. The end points are normalized to $1$.
\label{fig-rhob}}

\figure{Scaled one-particle density $\rho_\infty$ for the fermionic system
at half-filling, for the special values of $\lambda$. The curves for the
discrete and the continuous case are identical up to the accuracy of the
plot. The end points are normalized to $1$. Note that the curves for
$\lambda=\frac{1}{2}$ and $\lambda=2$ are identical, too.
\label{fig-rhof}}

\narrowtext
\begin{table}
\caption{Results for the critical exponents $\alpha$
from conformal finite-size scaling analysis and
numerical calculations for the special values of $\lambda$.
The estimated error for the last digit is given in parenthesis.
The $*$ indicates that the value of the corresponding exponent is
known exactly.}
\setdec 0.0000000000
\begin{tabular}{cccc}
          &              &\multicolumn{2}{c}{$\alpha$} \\
          &              & Conformal     & Numerical \\
          &$\lambda$     & Predictions   & Results \\
\tableline
          &$\frac{1}{2} $&$-\frac{3}{4} $& \dec-0.750002(5) \\
Bose      &$1^{*}       $&$-\frac{1}{2} $& \dec-0.500000(1) \\
          &$2^{*}       $&$ 0           $& \dec 0.0000000(1) \\
\tableline
          &$\frac{1}{2} $&$ \frac{1}{4} $& \dec 0.25001(4) \\
Fermi     &$1^{*}       $&$ 0           $& \dec 0.0000000(1) \\
          &$2           $&$ \frac{1}{4} $& \dec 0.25000(1) \\
\end{tabular}
\label{tab-res}
\end{table}
\bigskip

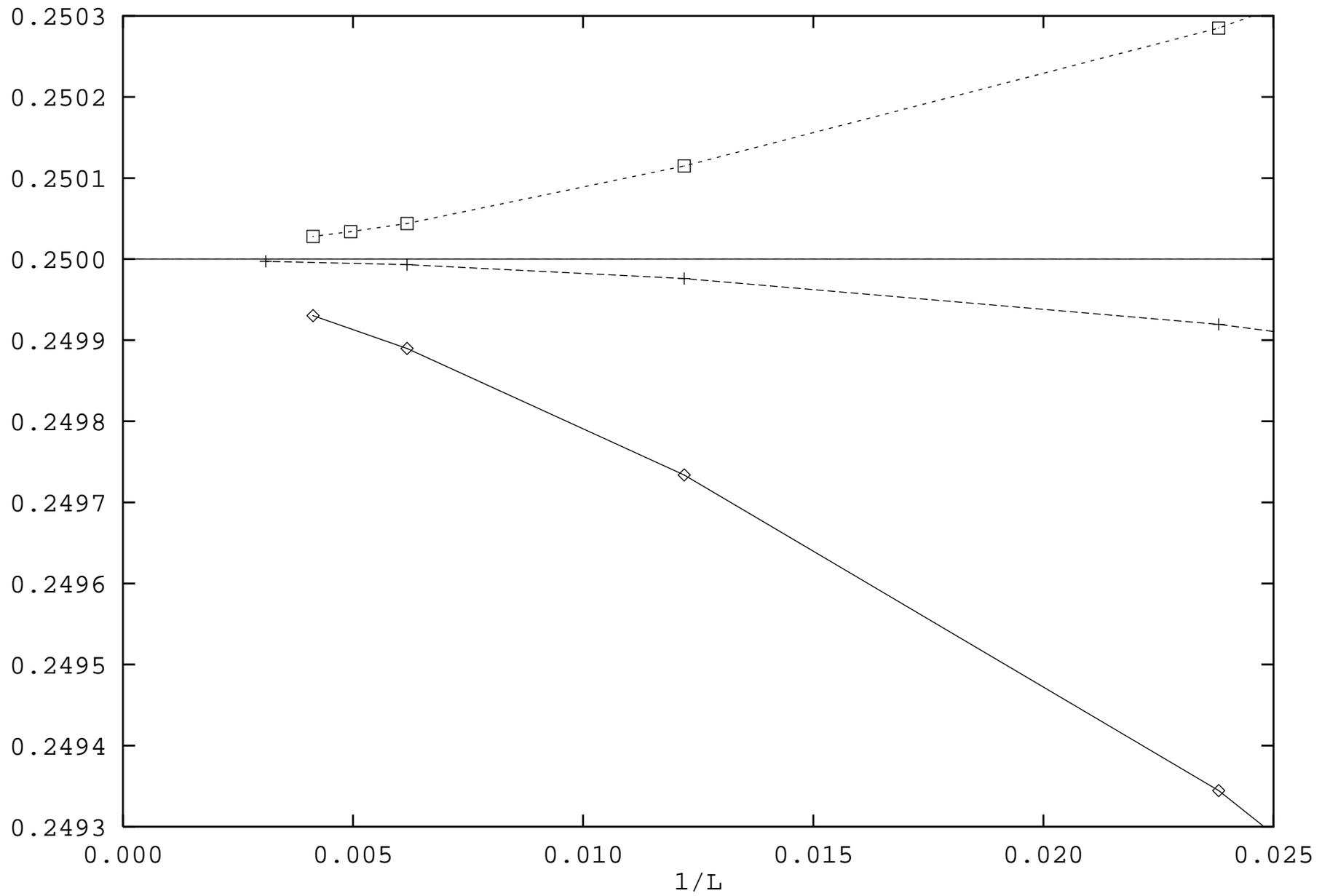
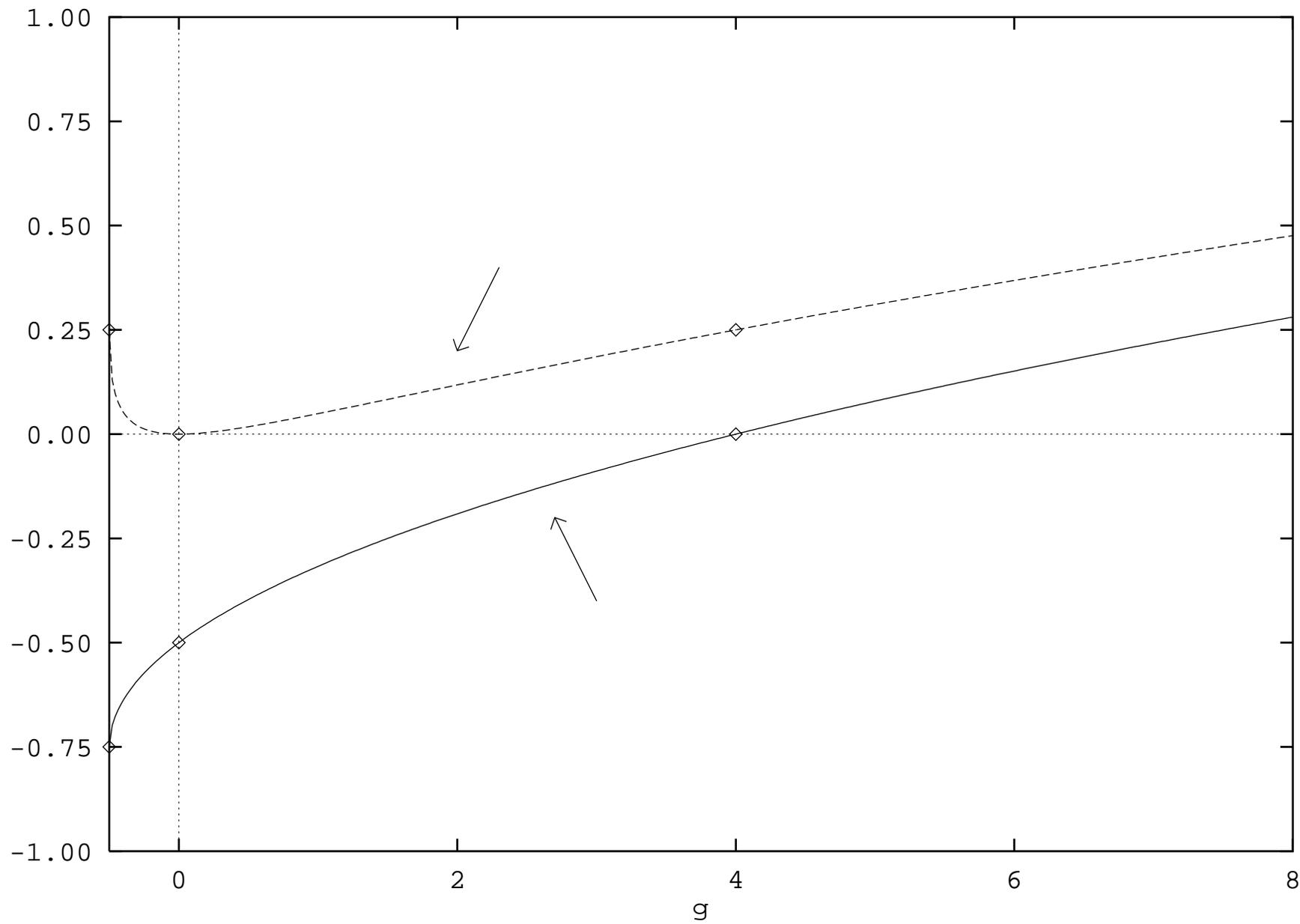
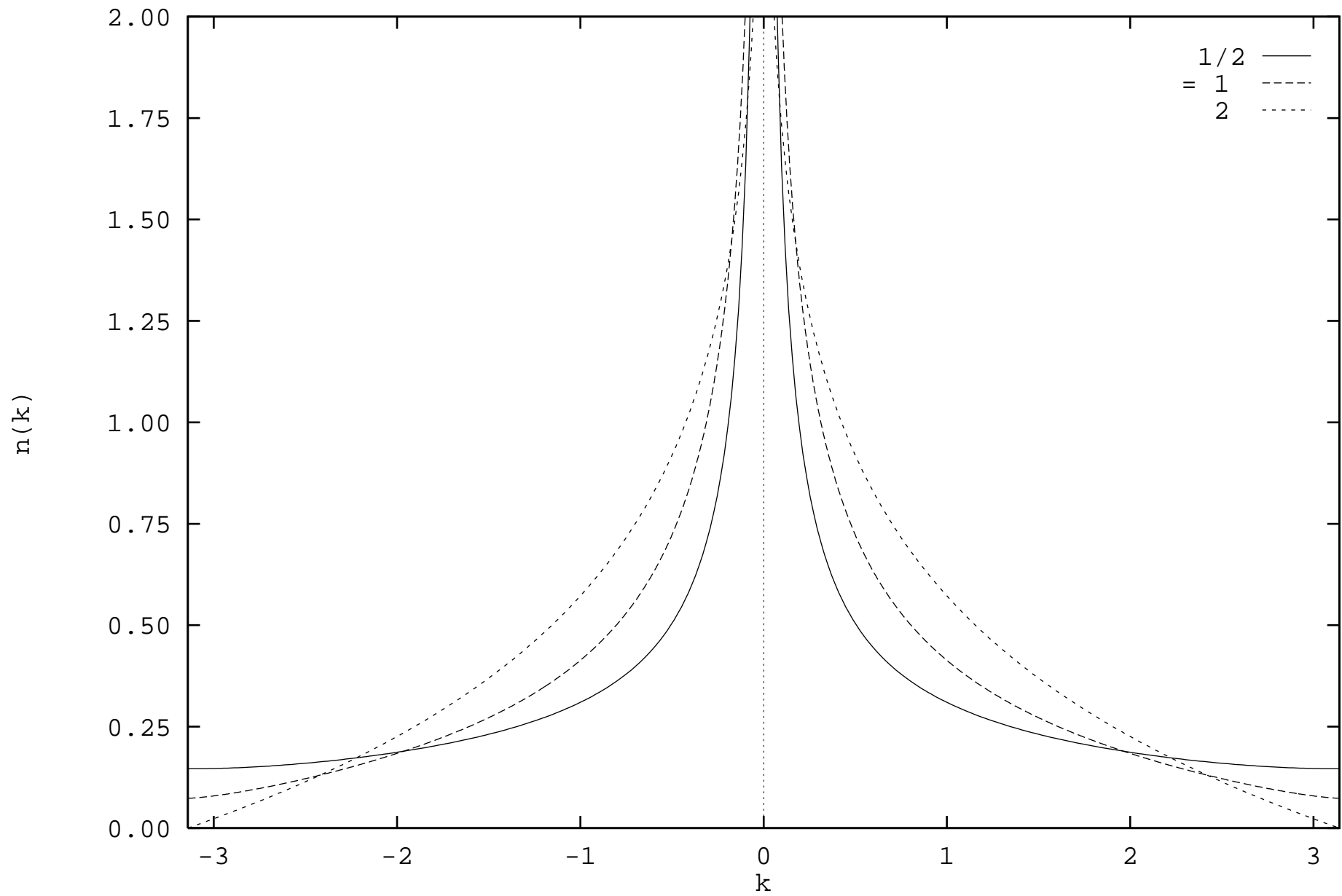
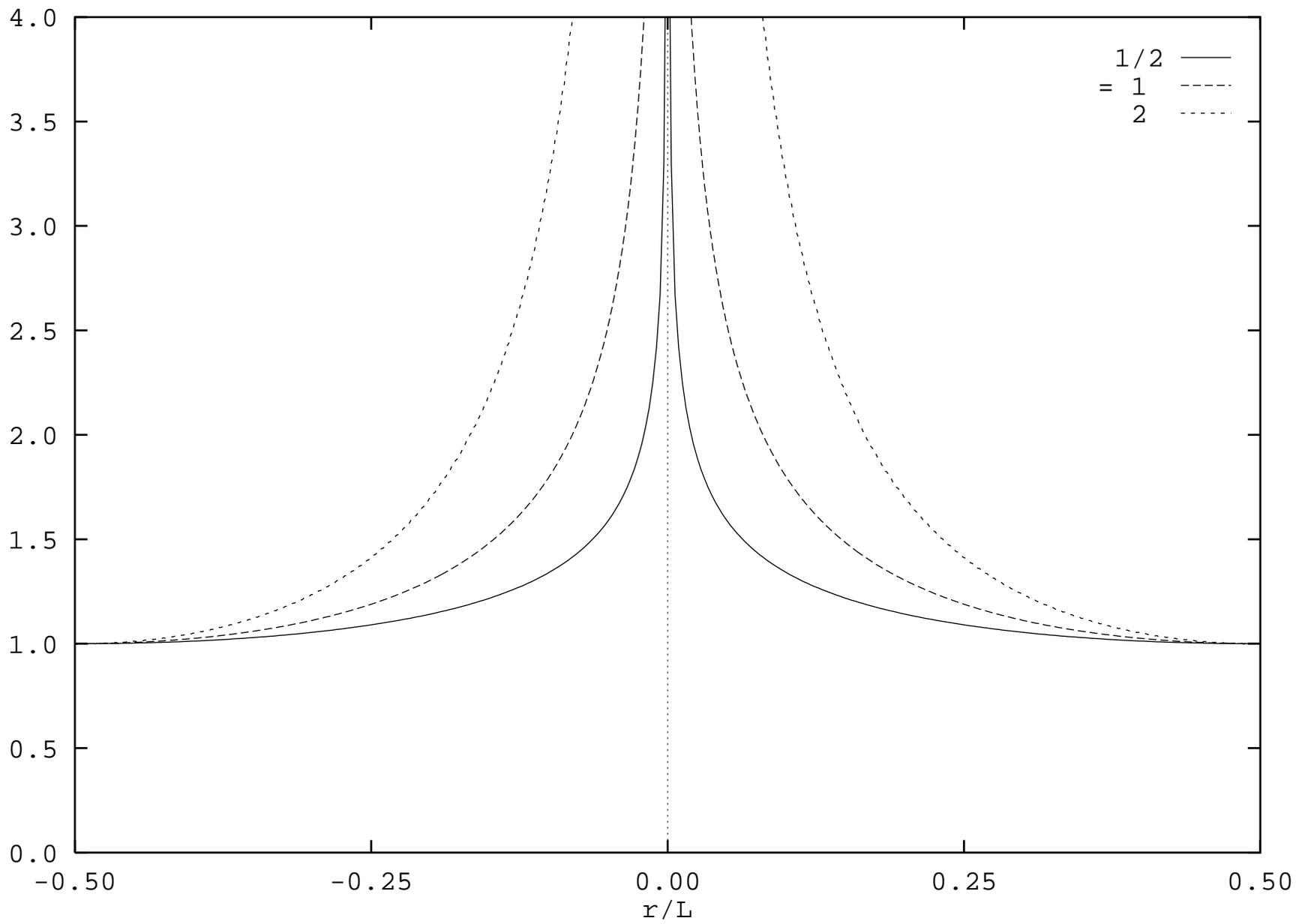
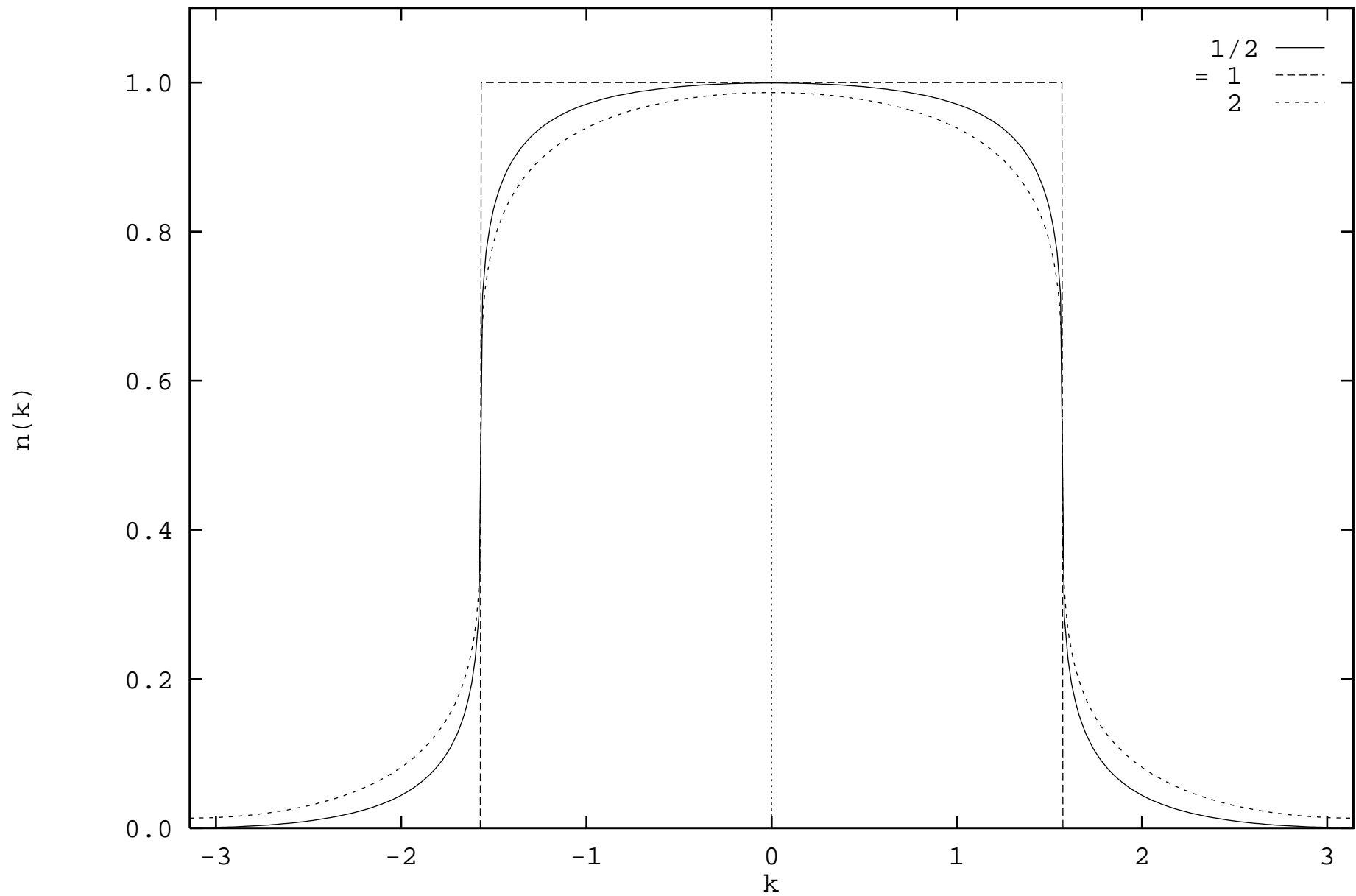
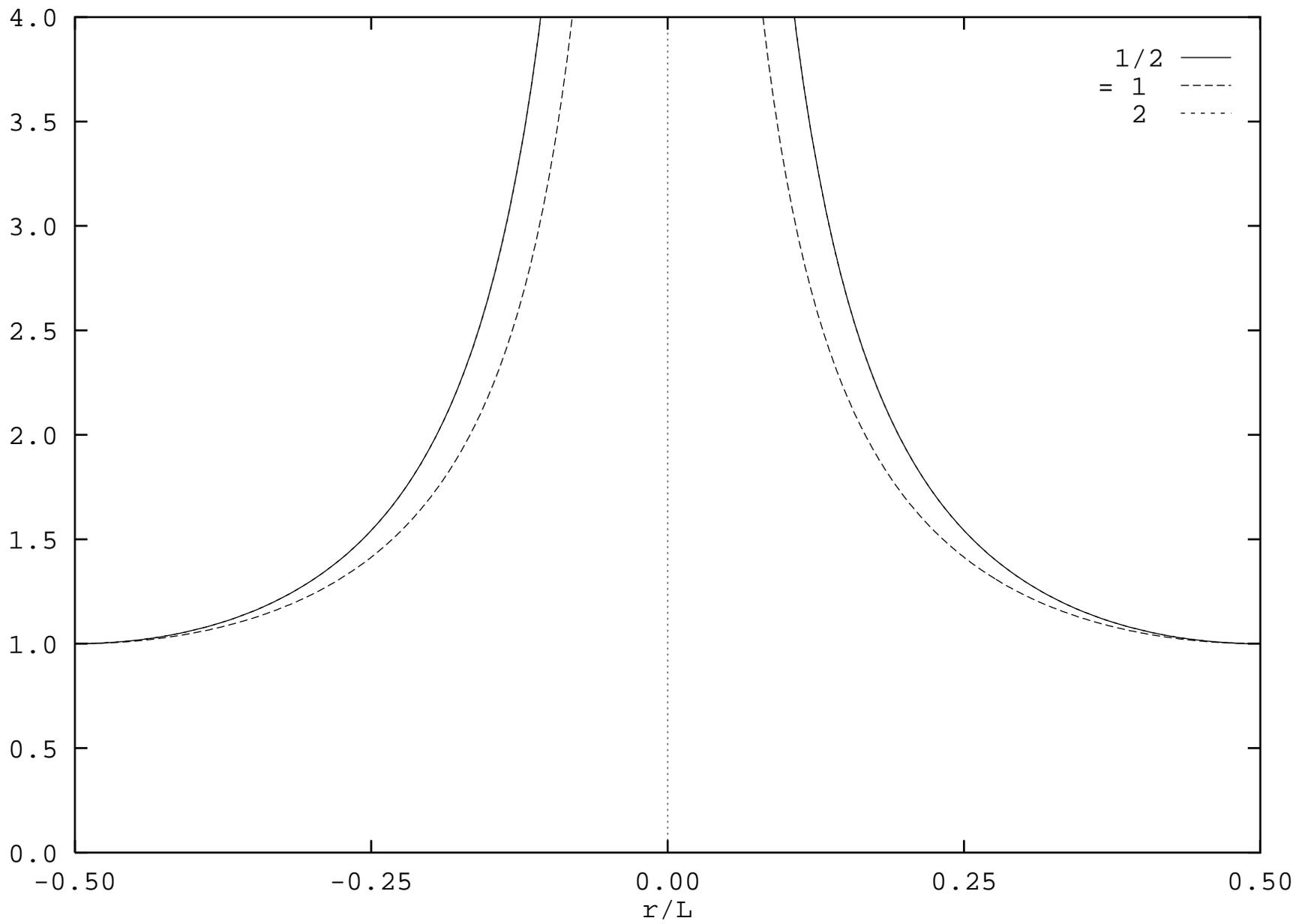
\end{document}